\documentstyle[twoside,fleqn,espcrc2]{article}
\def\beq{\begin{equation}}
\def\eeq{\end{equation}}
\def\bea{\begin{eqnarray}}
\def\eea{\end{eqnarray}}
\def\MPL{{\it Mod. Phys. Lett.}}
\def\NP{{\it Nucl. Phys.} }
\def\PL{{\it Phys. Lett.} }

\def\a{\alpha ^{\prime}}
\def\b{\tilde b}
\def\g{\tilde g}
\def\F{\tilde F}
\def\h{\tilde h}
\def\J{\tilde J}
\def\H{\tilde H}
\def\BI{\mbox{\tiny BI}}

\newcommand{\AmS}{{\protect\the\textfont2
  A\kern-.1667em\lower.5ex\hbox{M}\kern-.125emS}}

\title{Remarks on T-duality for open strings}

\author{H. Dorn$^{\mbox{\scriptsize a}}$ and H.-J. Otto\address{Humboldt-Universit\"at zu Berlin, 
Institut f\"ur Physik, Invalidenstr. 110, D-10115 Berlin}}
               
\begin{document}
\begin{titlepage}
\noindent
February 1997
\begin{flushright} HUB-EP-97/2\\ hepth@xxx/9702018
\end{flushright}
\vfill
\begin{center}
{\large\bf Remarks on T-duality for open strings${}^\dagger$}\\
\vskip 7.mm
{H. Dorn, H.-J. Otto }\\
\vskip 0.1cm                                                      
{\em Institut f\"ur Physik} \\
{\em Humboldt-Universit\"at Berlin}\\
{\em Invalidenstr.110, D-10115 Berlin} 
\end{center}
\vfill
\begin{center}
{\bf ABSTRACT}
\end{center}
\begin{quote}
This contribution gives in $\sigma $-model language a short review of recent 
work on T-duality for open strings in the presence of   
abelian or non-abelian gauge fields. Furthermore, it adds a critical
discussion of the relation between RG $\beta $-functions and the Born-Infeld
action in the case of a string coupled to a D-brane.
\vfill      \hrule width 5.cm
\vskip 2.mm
{\small\small
\noindent $^\dagger$Talk given at the $30^{\rm th}$ 
International Symposium Ahrenshoop on the Theory of Elementary 
Particles, Buckow, August 27 - 31, 1996; to appear in Nuclear 
Physics B (Proc. Suppl.).}   
\end{quote}
\end{titlepage}
\begin{abstract}
This contribution gives in $\sigma $-model language a short review of recent 
work on T-duality for open strings in the presence of   
abelian or non-abelian gauge fields. Furthermore, it adds a critical
discussion of the relation between RG $\beta $-functions and the Born-Infeld
action in the case of a string coupled to a D-brane.
\end{abstract}

\maketitle

\section{Introduction}
Dirichlet branes, i.e. hypersurfaces which the endpoints of strings
are confined to, play a fundamental role in the recent developments of the
string theory duality pattern \cite{pol,polwit,polrev} and the connected 
genesis of a unique underlying theory. Such D-branes arise from open 
strings with free ends by T-dualizing type I string theory 
or have to be added to type II theories in order to fulfil all the
duality requirements.

In $\sigma $-model language the T-duality transformation for open strings 
has been studied recently, both via formal functional integral 
manipulations as well as via canonical transformations 
\cite{alv,do,munich,bolo}. 

The aim of this contribution is twofold. At first we present in section 2 a 
streamlined discussion of T-duality for $\sigma $-models on two-dimensional
manifolds with boundary based on the functional integral formulation.
It is suited for a parallel study of both abelian and non-abelian
gauge fields coupling to the boundary of the manifold \cite{do,hd}.
The formalism can be used to give meaning to the notion of a matrix valued
D-brane position.

Secondly, we comment on a simple observation relevant for the issue of 
quantum equivalence of T-dual $\sigma $-models on manifolds with boundary.
To give to the results gained by formal manipulations with the functional
integral or by classical canonical transformations an exact status, one has to
perform a careful analysis of the renormalization properties. For
$\sigma $-models without boundary it has been shown that beyond 1-loop
order the Buscher rules \cite{bu} require modifications \cite{ts91,buda}.
Concerning the gauge field sector common folklore seems to take the absence
of such corrections for granted. This is based on the remarkable duality
property of Born-Infeld actions \cite{bats,alv}. In section 3 we will
point out that in the D-brane case the relation between the functional 
derivatives of the Born-Infeld action and the $\beta $-functions is an
inhomogeneous one. This is in collision with \cite{leigh} and implies the need 
for a more detailed study
of the problem in the gauge field sector, too. 
\section{Functional integral formulation of \\T-duality}
Our starting point is a generalized $\sigma $-model describing an open string 
moving in a target space with fields $\Psi =(G,B,\Phi)$ and $A$. The fields
$\Psi $ correspond to the massless excitations of closed strings. The
gauge field $A$ taking its values in the Lie algebra of some gauge group
${\cal G}$ corresponds to the massless excitations of open strings.
To identify the dual target space fields one can restrict oneself
to the partition function (A more complete study of the 2D field theory
would require the inclusion of source terms for $X$.)
\beq
Z[\Psi ,A]=\int DX^{\mu}e^{iS[\Psi ;X]}~\mbox{tr}Pe^{i\int _
{\partial M}A_{\mu}dX^{\mu}},
\label{1}
\eeq
with
\bea
\lefteqn{S[\Psi ;X]=S_M[\Psi ;X]-\frac{1}{2\pi}\int_{\partial M} 
k(s)\Phi ~ds~,}\nonumber \\
\lefteqn{S_M[\Psi ;X]=\frac{1}{4\pi \alpha ^{\prime}}\int _M d^2 z \sqrt{-g}}
\label{2}\\
&&~~~~~~~~\cdot \{ \partial _m X^{\mu} \partial _n X^{\nu} E^{mn}_{\mu \nu}
(X(z))\nonumber\\&&~~~~~~~~~~~~~~~~~~~~~~~~~~~+\alpha ^{\prime} R^{(2)}
\Phi (X(z))\},\nonumber\\[2mm]
&&E^{mn}_{\mu \nu}(X)=g^{mn}G_{\mu \nu}(X)+\frac{\epsilon ^{mn}}{\sqrt{-g}}
B_{\mu \nu}(X)~.
\nonumber
\eea
$R^{(2)}$ is the 2D curvature scalar on the manifold $M$, $k(s)$ the geodesic
curvature of the boundary $\partial M$.

In the case of a non-abelian gauge group ${\cal G}$, due to the path ordering
implied in the Wilson loop, the integrand of the functional integral
appears not in the standard form as an exponential of a local action.
To handle this complication we introduce an auxiliary field
$\zeta (s),~\bar{\zeta}(s)$ living on the one-dimensional space of the
variable $s$ parametrizing the boundary $\partial M$ of
the string world sheet $M$. We denote its free action by $S_0$. It has the propagator
\beq
\langle \bar{\zeta}_a (s_1)\zeta _b (s_2)\rangle _0=\delta _{ab}
\Theta (s_2-s_1)
\label{3}
\eeq
and  couples to $X^{\mu}$ via the interaction term
\beq
i\bar{\zeta}_a A_{\mu}^{ab}(X(z(s)))\zeta _b (s)\cdot
\partial _m X^{\mu}(z(s))\cdot\dot{z}^m(s)~.
\label{4}
\eeq
Then we can write (choosing $0\leq s\leq 1$)
\bea
\lefteqn{Z=\int DX^{\mu}~D\bar{\zeta}~D\zeta ~\bar{\zeta}_a(0)\zeta _a(1)}&&
\nonumber\\&&~~~~~~~~~~\cdot ~
e^{iS_0[\bar{\zeta}, \zeta ]}\exp (i\hat S[\Psi ,\bar{\zeta}A
\zeta ; X])~,
\label{5}
\eea
with
\bea
\hat S[\Psi ,C;X]&=&S[\Psi ;X]\label{6}\\
&+&\int_{\partial M} C_{\mu}(X(z(s)),s)\dot X^{\mu}~ds.
\nonumber
\eea
The field 
\beq
C_{\mu}(X(z),s)=\bar{\zeta}(s)A_{\mu}(X(z))\zeta (s)
\label{6a}
\eeq
in contrast to $A_{\mu}$ is no longer matrix valued. The case of an abelian 
gauge field is included in the formalism. However, then we can of course 
also drop the
auxiliary field and put $C=A$. In the following we interchange the order
of the $X$ and $\zeta $ integration and adapt for our open string case
the standard manipulations of the $X$-functional integral used to derive
the T-duality rules in the closed string case \cite{bu}. Apart from the final 
auxiliary 
field integration the only difference between abelian and non-abelian $A$
is the absence or presence of explicit $s$-dependence in $C$.

We now assume the existence of one Killing vector field $k^{\mu}(x)$ and 
invariance of our partition function under a diffeomorphism in the direction
of $k$
\beq
{\cal L}_k\Psi~=~0,~~~~~~~{\cal L}_kA_{\mu}~=~D_{\mu}a~.
\label{7}
\eeq
Choosing suitable adapted coordinates we have
\bea
\lefteqn{X^{\mu}~=~(X^0,X^M),}\label{8}\\
&&k^{\mu}~=~(1,0),~~~{\cal L}_k~=~\partial _0~.
\nonumber
\eea
Starting from an arbitrary gauge field fulfilling (\ref{7}) we can arrive at
$\partial _0A_{\mu}=0$ by a gauge transformation. Then all our target space
fields are independent of $X^0$
\beq
\partial _0\Psi~=~0,~~~~\partial _0A_{\mu}~=~0~.
\label{9}
\eeq
In the rest of the paper we take flat 2D metrics and a flat boundary
and disregard the dilaton field $\Phi $. Its transformation under T-duality
should arise from some functional Jacobian as in the closed string case
\cite{bu}.

To perform the dualization procedure we call the result of the $X$-integration 
in (\ref{5}) $\hat Z [\Psi ,C]$ and rewrite it with the help of a Lagrange 
multiplier field $\tilde X ^0$ as 
\bea
\lefteqn{\hat Z [\Psi ,C]~=~\int DX^{\mu}e^{i\hat S[\Psi ,C;X]}}\label{10}\\
&&=~
\int DX^{M}Dy_mD\tilde{X}^0 e^{i\bar{S}[\Psi ,C;X^M,y_m,\tilde{X}^0]}~,
\nonumber
\eea
with
\bea
\lefteqn{\bar S [\Psi ,C;X^M,y_m,\tilde{X}^0]=\frac{1}{4\pi \a}\int _M 
d^2 z}\nonumber\\
&&~~\{ \partial _m X^{M} \partial _n X^{N} E^{mn}_{MN}+
2\partial _mX^M y_n E^{mn}_{M0}\nonumber\\
&&~~~~~~~~~~~~~~~+y_m y_n E^{mn}_{00}+2\tilde{X}^0\epsilon ^{mn}
\partial _m y_n\}\nonumber\\
&&+\int _{\partial M} (C_N \partial _n X^N+C_0y_n)\dot{z}^n ds~.
\label{11}
\eea 
The dualized version will be obtained by performing the $y_m $ functional integral. After a shift of the integration variable suited to decouple the integral
in the closed string case we get
\bea
\lefteqn{\bar S=\frac{1}{4\pi \alpha ^{\prime}}\int _M d^2 z\{g^{mn}G_{00}
v_mv_n}\label{12}\\
&~~~~~~+E^{mn}_{MN}\partial _m X^{M} \partial _n X^{N}-\frac{g^{mn}}{G_{00}}
K_m K_n \}&\nonumber \\
&\int _{\partial M}\{C_N \partial _n X^N\dot z^n+v_m L^m-\frac{1}{G_{00}}K_m L^m
\}ds~,&
\nonumber
\eea
with
\bea
K^m=E^{mn}_{0N}\partial _n X^N+\epsilon ^{mn}\partial _n\tilde{X}^0~,
\nonumber\\
L^m=C_0\dot{z}^m+\frac{1}{2\pi
\alpha ^{\prime}}\tilde X^0\dot{z}^m~,\nonumber\\
v_m=y_m+\frac{K_m}{G_{00}}~.
\label{13}
\eea
Due to the dependence of $L^m$ on $X^M,~\tilde X ^0$ the $v$ integral even
after the rescaling $v\rightarrow \sqrt{G_{00}}v$ does not decouple.

Since the $v$ integral is ultralocal we can use
\bea
\lefteqn{\int _M Dv_me^{i\bar S}=\int _M Dv_me^{ \left (i\bar S-i\int _{\partial M}
v_mL^mds\right )}}\nonumber\\
&&~~~~~~~~\cdot \int _{\partial M}Dv_me^{ \left (i\int _{\partial M}
v_mL^mds\right )}~.
\label{14}
\eea
Then the $v_m$ integral on the boundary $\partial M$ produces a functional
$\delta$ function which imposes for the remaining functional integral the
constraints $L^m(z(s))=0,~~m=1,2$, i.e.
\beq
2\pi \alpha ^{\prime}C_0 (X^M(z(s)),s)~+~\tilde X^0(z(s))=0~.
\label{15}
\eeq
Including the remaining bulk $v_m$ integral into the overall normalization
we finally arrive at (The $\ast $ below the integration symbol indicates
the boundary condition (\ref{15}).)
\beq
\hat Z[\Psi ,C]~=~\int _{\ast}D \tilde X^{\mu}\exp (i\hat S[\tilde
{\Psi},\tilde C ;\tilde X])~,
\label{16}
\eeq
with (note $\tilde X^M=X^M$)
\beq
\tilde X^{\mu}~=~(\tilde X^0,X^M)~,
\label{17}
\eeq
\beq
\tilde A_{\mu}~=~(0,A_M)~,~~~~~~\tilde C_{\mu}~=~\bar{\zeta}\tilde A_{\mu}\zeta ~,
\label{18}
\eeq
\bea
\lefteqn{\tilde G_{00}=\frac{1}{G_{00}}~,~~~~~\tilde G_{0M}=\frac{B_{0M}}{G_{00}}~,}
\nonumber\\
\lefteqn{\tilde G_{MN}=G_{MN}-\frac{G_{M0}G_{0N}+B_{M0}B_{0N}}{G_{00}}~,}
\label{19}
\eea
\bea
\lefteqn{\tilde B_{0M}=\frac{G_{0M}}{G_{00}}~,}\nonumber\\
\lefteqn{\tilde B_{MN}=B_{MN}-\frac{G_{M0}B_{0N}+B_{M0}G_{0N}}{G_{00}}~.}
\label{20}
\eea
For later use we denote $\hat Z[\Psi ,C]$ understood as a functional of
$\tilde {\Psi},~\tilde C$ and the function $-2\pi\a C_0$ determining the boundary condition in (\ref{16}) by $\cal F$
\beq
\hat Z[\Psi ,C]~=~{\cal F}[\tilde{\Psi},\tilde C\vert-2\pi\a C_0]~.
\label{21}
\eeq 

The formulas (\ref{19}, \ref{20}) coincide with those in the closed string case
\cite{bu}. The gauge field $A^0(X^M)$ of the original model determines via
(\ref{15}) a Dirichlet boundary condition for the functional integrand
in the integration defining the dual model (\ref{16}). What concerns the 
quantized theory, the Dirichlet condition, therefore, has to be considered
as an external constraint. In contrast, on the classical level there are always
equivalent formulations for a boundary condition, either as an external
constraint or as part of the stationarity condition of the action, possibly 
after a suitable addition of some boundary interaction. A discussion of 
T-duality and classical canonical transformations has been given in 
\cite{do,bolo} extending corresponding results for the closed string case 
\cite{alvg}.

In the abelian case we can drop the $\zeta $'s and put $C=A$, $Z[\Psi ,A]=
\hat Z[\Psi ,A]$. The boundary condition (\ref{15}) has to be interpreted as 
a constraint confining the end points of the string to the hypersurface
(Dirichlet brane)
$$\tilde X^0=-2\pi \a A_0(\tilde X^M)$$
with free movement on the brane.

In the nonabelian case (\ref{16}) still has a D-brane interpretation.
There is a one parameter family of D-branes
$$\tilde X^0=-2\pi \a\bar\zeta (s)A_0(\tilde X^M)\zeta (s)~.$$
The endpoint of the string at string world sheet boundary parameter value $s$
has to sit on the D-brane for parameter value $s$.

The final $\zeta $-integration results in ordering the matrices sandwiched between
$\bar{\zeta}(s)$ and $\zeta (s)$ with respect to increasing $s$. But after
performing the functional integral over the world surfaces $X^{\mu}(z)$
there is no longer any correlation between a given target space point and
$s$. The situation is more transparent if one treats (\ref{5}) and (\ref{10})
within the background field method ($bm$). Then both in ${\cal F}_{bm}$
and $Z_{bm}$ all dependence on target space coordinates is realized via
a classical string world sheet configuration $X_{cl}$. The result of the 
$\zeta $-integration is then \cite{do}
\bea
\lefteqn{Z_{bm}[\Psi ,A;X_{cl}]=}\nonumber\\
&&\mbox{tr}P{\cal F}_{bm} [\tilde{\Psi}, \tilde{A};
\tilde X_{cl}\vert -2\pi \alpha ^{\prime}A_0]~.
\label{22}
\eea
Path ordering now refers to the classical path $\tilde X_{cl}(z(s))$ and
involves both the matrices appearing in the second argument
of ${\cal F}_{bm}$ as well as $A_0$ entering via the argument specifying the
boundary condition.

The insertion of a matrix as boundary condition is performed {\it after}
${\cal F}_{bm}$ has been calculated with scalar (not matrix valued) 
$s$-dependent boundary condition. In this formalism we can avoid wondering 
about the target space interpretation of matrix valued boundaries. The 
situation is similar to dimensional regularization, the change from integer 
$n$ to complex $n$ is performed {\it after} $\int d^n x ...$ has been 
calculated for integer $n$.
\section{An observation concerning quantum equivalence of T-dual models
for open strings}
The manipulations with the functional integral reviewed in the previous 
section are of pure formal nature. No attention has been paid both to 
functional Jacobians and renormalization issues. 

In the closed string case, 
studied in more detail so far, there are thorough arguments for the quantum 
equivalence of conformal theories T-dual to one another \cite{rove}. Away from
the points of conformal invariance the situation is less clear. The Buscher
formulae (\ref{19}), (\ref{20}) can be derived by pure classical 
considerations on the basis of canonical transformations \cite{alvg}.
The Weyl anomaly coefficients, related closely to the renormalization group
$\beta $-functions \cite{shtsos}, fulfil at one-loop order the consistency 
equations derived from requiring (\ref{19}) and (\ref{20}) for the 
renormalized target space fields \cite{ts91,ha}. However, insisting on 
quantum equivalence in higher orders of renormalized perturbation theory
requires a modification of the Buscher rules starting at two loops 
\cite{ts91,buda}.
\footnote{Repeating the formal functional integral manipulations with 
regularized expressions (still ignoring functional Jacobians) one expects 
the unmodified rules (\ref{19}),(\ref{20}) to hold for the bare target space 
fields.}

We now want to comment on some simple observations concerning quantum 
equivalence of the models related by (\ref{17})-(\ref{20}) in the case
of open strings. Then all the problems present in the absence of a boundary 
of the world sheet are present, too. In addition the Dirichlet boundary 
condition should refer at first sight to the bare quantities.
For shortness we restrict ourselves to the abelian case and discuss the gauge 
field $\beta $-function only.

In leading order of an expansion in the derivatives of the field strength
$F$ this $\beta $-function is \cite{cal} (We absorb a factor $2\pi\a$ into $F$.
$H$ denotes the field strength for $B$.)
\bea
\beta^{(A)}_{\mu}&=&(B+F)_{\mu}^{~\nu}\partial _{\nu} \Phi
\label{23}\\
&+&J^{\alpha\beta}\{ D_{\alpha}(B+F)_{\beta\mu}\nonumber\\
&&~~~~~~+\frac{1}{2}(B+F)_{\mu}^{~\nu}H_{\nu\alpha\gamma}
(B+F)^{\gamma}_{~\beta}\}\nonumber
\eea
with
\beq
J^{\alpha\beta}~=~\left ((G-(B+F)^2)^{-1}\right )^{\alpha\beta}~.
\label{24}
\eeq
The vanishing of this $\beta $-function is equivalent to the stationarity 
condition of the generalized Born-Infeld action \cite{cal}
\bea
\lefteqn{S_{\BI}[\Phi ,G,B,F]~=}&&\label{25}\\
&&~~~~~~~~\int d^Dx~e^{-\Phi}\sqrt{\det (G+B+F)}~,
\nonumber
\eea
\beq
\frac{\delta S_{\BI}}{\delta A_{\mu}}=-e^{-\Phi}\sqrt{\det(G+B+F)}
J^{\mu\nu}\cdot\beta^{(A)}_{\nu}~.
\label{26}
\eeq
The action $S_{\BI}$ has a remarkable duality property \cite{bats,alv}.
For target space fields of the type discussed in the previous section
($\partial _0\Psi =\partial _0A=0$) one finds by simple manipulations,
based on the identity ($M_{\mu\nu}$ arbitrary matrix, $M_{00}\neq 0$)
$$ \det (M_{\alpha\beta})=M_{00}\det \left (M_{AB}-\frac{M_{A0}M_{0B}}{M_{00}}
\right ),$$
the relation
\bea
\lefteqn{e^{-\Phi}\sqrt{\det (G+B+F)_{\mu\nu}}~=}\label{27}\\
&&~~~~~~~~~~~~e^{-\tilde{\Phi}}\sqrt{\det (\tilde g+\tilde b+\tilde F)_{MN}}~.
\nonumber
\eea
Here $\tilde g$ and $\tilde b$ are the metric
and the antisymmetric tensor induced on the D-brane defined by
$$f^0(X^M)=-A_0(X^M)~,~~~f^M=X^M~, $$
i.e.
\beq
\tilde g _{MN}~=~\tilde G_{\mu\nu}\partial _Mf^{\mu}\partial _Nf^{\nu}
\label{28}
\eeq
and similar for $\tilde b,~\tilde B$. The fields $\tilde G,~\tilde B,~\tilde A$ are given by (\ref{18})-(\ref{20}), $\tilde{\Phi}=\Phi -\frac{1}{2}\log G_{00}.$

The $\beta $-functions of a $\sigma $-model describing a string with its
endpoints confined to some Dirichlet brane have been calculated in 
ref.\cite{leigh}. Furthermore, in this paper it has been argued that the vanishing
of the $\beta $-functions for the gauge field on the D-brane as well as that 
for the D-brane position is equivalent to the stationarity condition
of the Born-Infeld action restricted to the D-brane $S_{\BI}[\tilde{\Phi},\tilde g,\tilde b,\tilde F]$. If this last statement was correct, due to (\ref{27})
the gauge field part in the issue of quantum equivalence of T-dual models
would be completely settled. However, while we agree with the result of the 
$\beta $-function calculation of \cite{leigh} (see also \cite{hd}),
the relation of these $\beta $-functions to the functional derivatives
of the D-brane BI-action needs some more discussion.\\

The $\beta $-functions just mentioned are
that for the gauge field $\tilde A_M$ living on the D-brane
\bea
\lefteqn{\tilde{\beta}^{(\tilde A)}_M~=~(\b+\F)_M^{~~N}\partial _N\tilde{\Phi}}&&\label{29}\\
&&~~~+~\J^{NL}\{D_L(\b+\F)_{NM}\nonumber\\
&&~~~~~~~~~~~~~+\frac{1}{2}(\b+\F)_M^{~~C}\h_{CNE}(\b+\F)^E_{~L}
\nonumber\\
&&~~~~~~~~~~~~~~~~~~~+~(\b+\F)_M^{~~C}\partial _Cf^{\mu}K_{\mu NL}\}
\nonumber
\eea   
and that for the D-brane position $f_{\mu}$
\bea
\tilde{\beta}^{(f)}_{\mu}&=&-\partial _{\mu}\tilde{\Phi} \label{30}\\
&+&\J^{NL}\{\frac{1}{2}(\b+\F)_L^{~~C}\H_{\mu NC}-K_{\mu NL}\},
\nonumber  
\eea
with $\J$ related to $\g,~\b,~\F$ like $J$ to $G,~B,~F$ in (\ref{24})
and the curvature of the D-brane given by
\beq
K^{\mu}_{~~ NL}=D_N\partial _Lf^{\mu}~.
\label{31}
\eeq
The two $\beta $-functions fulfil the equation
\bea
\tilde{\beta}^{(\tilde A)}_M&=&\J^{NL}D_L(\b+\F)_{NM}\label{32}\\
&-&(\b+\F)_M^{~~N}
\partial _{N}f^{\mu}\cdot \tilde{\beta}^{(f)}_{\mu}~.
\nonumber
\eea
For the functional derivatives of $S[\tilde{\Phi},\g,\b,\F]$ with
respect to $\tilde A$ and $f$ (The last one for variations
orthogonal to the D-brane.) we find
\bea
\frac{\delta S_{\BI}}{\delta \tilde A_M}&=&-e^{\tilde{\Phi}}
\sqrt{\det (\g+\b+\F)}\label{33}\\
&\cdot &\left (\tilde{\beta}^{(\tilde A)}_M-\J^{NL}(\b+\F)_M^{~~C}\partial
_Cf^{\mu}K_{\mu NL}\right )~,\nonumber
\eea
\bea
\frac{\delta S_{\BI}}{\delta f^{\mu}_{\perp}}&=&e^{\tilde
{\Phi}}\sqrt{\det (\g+\b+\F)}\cdot \tilde{\beta}^{(f)}_{\mu}\label{34}\\
&-&2\tilde B_{\mu\nu}\partial _Nf^{\nu}\cdot \frac{\delta S_{\BI}}{\delta \tilde A_N}~.\nonumber
\eea
The last two formulas disagree with ref.\cite{leigh} by the second terms on
their r.h.s. We want to stress that at least (\ref{33}) is obvious without
any calculation. The subtraction of the curvature term in (\ref{33}) ensures
that the gauge field derivative has the same form as in (\ref{23}). This
has to be correct since the gauge field $\tilde A_M$ is defined on the D-brane
only. In contrast to $\g$ and $\b$ it is not induced from target space fields
and does not ``feel'' the embedding.

Now the problem connected with (\ref{33}),(\ref{34}) is due to the presence of
an inhomogeneity (the curvature term) in the otherwise linear relation
between the $\beta $-functions and the functional derivatives of the 
BI-action. We no longer can claim the equivalence of the vanishing
of the $\beta $-functions and the stationarity of 
$S_{\BI}$.

As far as one is interested in the equivalence of BI stationarity and the 
conformal invariance condition of the D-brane model there is still a way out.
The Weyl anomaly coefficients are given by the $\beta $-functions plus
some additive terms \cite{shtsos,bdos} usually contributing in higher orders 
only. While the $\beta $'s are not 
diffeomorphism invariant the Weyl anomaly coefficients are invariant
under diffeomorphisms. Hence as long as these additive terms are not 
calculated one should state conformal invariance as the vanishing of the
$\beta $-functions up to terms which can be generated by a diffeomorphism.
The disturbing term in (\ref{33}) has the form of a product of $(\b+\F)$
times a vector . Therefore, it can be generated by a combination of a diffeomorphism and a gauge transformation. Of course a final statement can be made only
after a calculation of the difference between the Weyl anomaly coefficients 
and the corresponding $\beta $-functions has been done.   
\\[5mm]
\noindent
{\bf Acknowledgement}\\
We would like to thank J. Borlaf, S. F\"orste, A. Kehagias, L. Palla, 
C. Preitschopf, C. Schmidhuber and A. Tseytlin for useful discussions.


\begin{thebibliography}{9}

\bibitem{pol}
J. Polchinski, {\it Phys. Rev. Lett.} {\bf 75} (1995) 4724, hep-th/9510017

\bibitem{polwit}
J. Polchinski, E. Witten, \NP {\bf B460} (1996) 525, hep-th/9510169 

\bibitem{polrev}
J. Polchinski, S. Chaudhuri, C.V. Johnson, preprint {\it Notes on D-Branes},
hep-th/9602052\\
J. Polchinski, preprint {\it TASI lectures on D-branes}, hep-th/9611050

\bibitem{alv}
E. Alvarez, J.L.F. Barbon, J. Borlaf, \NP {\bf B479} (1996) 218, 
hep-th/9603089

\bibitem{do}
H. Dorn, H.-J. Otto, \PL {\bf B381} (1996) 81, hep-th/9603186

\bibitem{munich}
S. Forste, A.A. Kehagias, S. Schwager, \NP {\bf B478} (1996) 141,
hep-th/9604013\\
S. Forste, A.A. Kehagias, S. Schwager, hep-th/9610062, hep-th/9611060

\bibitem{bolo}
J. Borlaf, Y. Lozano, {\it Aspects of T-duality in open strings}
preprint hep-th/9607051\\
Y. Lozano, {\it Duality and canonical transformations}
preprint hep-th/9610024

\bibitem{hd}
H. Dorn, {\it Nonabelian gauge field dynamics on matrix D-branes,} preprint
hep-th/9612120

\bibitem{bu}
T.H. Buscher, {\it Phys. Lett.} {\bf B194} (1987) 59,
              {\it Phys. Lett.} {\bf B201} (1988) 466

\bibitem{ts91}
A.A. Tseytlin, \MPL {\bf A6} (1991) 1721

\bibitem{buda} 
J. Balog, P. Forgacs, Z. Horvath, L. Palla, \PL {\bf B388} (1996) 121,
hep-th/9606187\\
J. Balog, P. Forgacs, Z. Horvath, L. Palla, {\it Nucl. Phys. B (Proc. Suppl.)}
{\bf 49} (1996) 16, hep-th/9601091\\
J. Balog, P. Forgacs, Z. Horvath, L. Palla, {\it Contribution to these
Proceedings}

\bibitem{bats}
C. Bachas, \PL {\bf B374} (1996) 37, hep-th/9511043\\
A.A. Tseytlin, \NP {\bf B469} (1996) 51, hep-th/9602064

\bibitem{leigh}
R.G. Leigh, {\it Mod. Phys. Lett.} {\bf A4} (1989) 2767

\bibitem{alvg}
E. Alvarez, L. Alvarez-Gaum\'e, Y. Lozano, {\it Nucl. Phys.} {\bf B41},
Proc. Suppl. (1995) 1, hep-th/9410237

\bibitem{rove}
M. Rocek, E. Verlinde, \NP {\bf B373} (1992) 630, hep-th/9110053

\bibitem{shtsos}
G.M. Shore, \NP {\bf B286} (1987) 349\\
A.A. Tseytlin, \NP {\bf B294} (1987) 383\\
H. Osborn, \NP {\bf B308} (1988) 629

\bibitem{ha}
P.E. Haagensen, \PL {\bf B382} (1996) 356, hep-th/9604136

\bibitem{cal}
C.G. Callan, C. Lovelace, C.R. Nappi, S.A. Yost, \NP {\bf B288} (1987) 525

\bibitem{bdos}
K. Behrndt, H. Dorn, {\it Int. Jour. Mod. Phys.} {\bf A7} (1992) 1375\\
H. Osborn, \NP {\bf B363} (1991) 486

\end{thebibliography}
\end{document}